%% file: main.tex
\documentclass[twocolumn]{openjournal}
\usepackage{amsmath}
\usepackage{xurl}
\usepackage[colorlinks=true,
linkcolor=blue,
citecolor=blue,
urlcolor=blue]{hyperref}
\usepackage{booktabs}
\usepackage{orcidlink}

\begin{document}
	\title{Radial Pulsations in Polaris: A Secondary Science Application of Cherenkov Telescopes via Intensity Interferometry}
	\date{}
	
	\author{Km Nitu Rai$^{1,2\orcidlink{0000-0002-9939-805X}}$
		Prasenjit Saha$^{3\orcidlink{0000-0003-0136-2153}}$
		and Subrata Sarangi$^{4,5\orcidlink{0000-0001-6500-9987}}$}
	
	\affiliation{$^{1}$School of Physics, Indian Institute of Science
		Education and Research Thiruvananthapuram, Maruthamala PO, Vithura,
		Thiruvananthapuram 695551, Kerala, India.}
	
	\affiliation{$^{2}$Aryabhatta Research Institute of Observational Sciences,
		 Manora Peak, Nainital 263129, India.}
	
	\affiliation{$^{3}$Physik-Institut, University of Zurich,
		Winterthurerstrasse 190, 8057 Zurich, Switzerland.}
	
	\affiliation{$^{4}$School of Applied Sciences, Centurion University of
		Technology and Management, Odisha-752050, India.}
	
	\affiliation{$^5$ Visiting Associate, Inter-University Centre for 
		Astronomy and Astrophysics, Post Bag 4, Ganeshkhind, Pune 411007, 
		Maharashtra, India.}
	\email{niturai201296@gmail.com}
	
	\begin{abstract}
		Ground-based Cherenkov telescopes, which are typically inoperative during moonlit nights for gamma-ray observations, offer a valuable opportunity during this time for secondary scientific applications through Intensity Interferometry (II). Recent developments and observations suggest that implementing II instrumentation on existing and planned Imaging Atmospheric Cherenkov Telescopes (IACTs) can significantly advance optical stellar measurements. Motivated by the resurgence of II efforts over the past two decades, this work presents simulations demonstrating the estimation of stellar parameters for a radially pulsating star, such as Polaris, using either a single telescope or multiple telescopes. For single-telescope simulations, we assume that the photon pixels in the camera are mapped onto four distinct regions of the aperture, generating multiple baselines and enabling enhanced observational plane coverage. These results highlight the potential of Cherenkov telescopes in India for high-resolution optical astronomy during otherwise inoperative periods and offer promising insights into the characterization of bright stellar objects with unprecedented precision.
	\end{abstract}
	
	\maketitle
	
	\input{Introduction}
	\input{polaris_observation}

	\input{mace}
	\input{tactic}
	\input{summary}

	\input{conclusion}
	\section*{Acknowledgments}
We sincerely thank the anonymous referee for the constructive comments and valuable suggestions, which have significantly improved the quality of this manuscript. One of the authors (SS) sincerely acknowledges the Inter-University Centre for Astronomy and Astrophysics (IUCAA), Pune, India, for providing computing facilities and hospitality through its Visiting Associate Programme during the preparation and finalization of this manuscript.
		
\bibliographystyle{aasjournal}
\bibliography{main.bib}

\end{document}

%% file: introduction.tex
\section{Introduction}
Intensity Interferometry (II), a technique to observe point sources in sky as extended objects using optical telescopes, was proposed and validated by Hanbury Brown and Twiss \citep{HBT56-2} wherein the authors reported the results of applying this technique to observe Sirius. This technique was subsequently also applied by Hanbury Brown and  team at the historic Narrabri Stellar Intensity Interferometer with the then-available photon detectors. By 1974, this team reported the measurement of the angular diameters of 32 stars in the solar neighborhood \citep{HB74}. However, due to the limitations imposed by the speed of the contemporary photon counters, II passed into a period of slumber. In recent years, with fast photon counters available, there has been a resurgence of interest in II. Several experiments and observations have been conducted using Imaging Atmospheric Cherenkov Telescopes (IACTs) as II telescopes, equipped with faster photon counters at the VERITAS, MAGIC, HESS and ASTRI arrays \citep{acharyya2024angular, abe2024performance, vogel2025simultaneous, 10.1093/mnras/stac1617}. This revival emphasizes a need for renewed efforts to explore the potential of II in India as a valuable technique to observe bright stellar objects in the northern hemisphere using IACTs. The similarity in the basic operational procedures of IACTs and the proposed II apparatus is noteworthy. Both of the processes involve photon collection (at the objectives of the telescopes) and counting of the photons received at the photon counters placed on the focal planes (as part of the ``eye-pieces") of the objectives. In both the cases, information on the source of the detected radiation is extracted from the correlation between photons received at the detectors (the photon counters). Then comes the distinction. During the moonlit nights, the operations of IACTs as Cherenkov Radiation detectors for observing the faint Cherenkov Showers produced by high energy cosmic $\gamma$-photons in the upper atmosphere are rendered challenging. However, II measurements can be carried out during such nights at least on observable brighter stars or the stars away from the moon's trajectory. This similarity and the distinctive operational possibility during moonlit nights have enabled the use of IACTs \citep[such as VERITAS][]{Kieda22} as II measurement facilities. Upcoming facilities like the ASTRI Mini-Array in Tenerife (Spain) also look forward to carrying out II observations \citep{astri2023} alongside observations of UHE/ VHE $\gamma$-ray events.

At the most basic level, II technique involves measurement of the intensity of radiation received from a stellar source (say, a single, binary or multiple star system) by a pair of telescopes separated by a fixed baseline. The intensity is measured by counting photons of a chosen wavelength $\lambda$ and within a narrow band width $\Delta \lambda$ received at the photon counters placed at the foci of the telescopes. The bandwidth $\Delta \lambda$ corresponds to a frequency bandwidth $\Delta \nu = \left|c\,\Delta \lambda/\lambda^2\right| $ and to a coherence time $\Delta \tau = 1/{\Delta\nu}$. The photons received at the photon counters within the coherence time $\Delta \tau$ are considered coherent and correlated to generate the interference patterns. One of the advantages of II is that these signals from the telescopes can be correlated either real time or be saved and correlated off-line later.

The fundamentals of II are rooted in the van Cittert \citep{Cittert34} - Zernike \citep{Zernike38} theorem. From a practical point of view, Hanbury Brown and Twiss extended their results \citep{HB54, HBT56-1} of measurements carried out on radio stars using radio-frequency detectors to the visible range by suggesting the replacement of a pair of aerials - rf detector - low pass filter sets with a pair of the mirror - photocell sets where the mirrors acted as the light collectors. In recent times, several laboratory experiments examining various aspects of II have been reported \citep{dravins2014stellar, dravins2015long, rai2025interference}. 

It is also worthwhile to note a list of merits and limitations of II over its alternative method, namely, the Michelson Stellar Interferometry (MSI). The major advantage stems from the fact that the strict requirement of coherence, as in MSI, is dispensed with. This allows longer baselines, enhancing thereby the angular resolution of the observation. IACT facilities like VERITAS and HESS which are currently used for II use telescopes with diameters $\sim$10 m and baselines of $\sim$100-200 m to achieve angular resolution beyond $\sim$ milli-arc-seconds range. The classical optical telescopes at La Palma and Chile envisage baselines of $\sim$2 km or more with potential for unprecedented angular resolutions upto $\sim$30 micro-arc-seconds \citep{IAC2024}. Besides, telescopes of comparatively low optical quality with cheaper optics are acceptable in II thus making establishment of such facilities highly cost-effective compared to the demanding Adaptive Optics of MSI telescopes. The other issue with ground-based optical observation is that of atmospheric turbulence and II effectively resolves this issue in the following manner: II measurements are based on the second-order coherence function $g^{(2)}(\tau)$, which tracks the intensity correlations over very short time-scales (nanoseconds). This time-scale is much smaller than the typical time-scales (of milliseconds to seconds) of atmospheric turbulences. As the atmospheric turbulence primarily affects the phase of the incoming wavefronts and not the intensities themselves, II measurements are less affected. As far as the issue of image reconstruction of the source is concerned, both MSI and II face difficulties due to incomplete $(u,v)$ plane coverage of the interferometric data and other noises and artefacts. In addition, II has the additional disadvantage of loss of phase information. In the case of MSI, the deconvolution methods with incomplete $(u,v)$ plane data and methods of noise suppression due to artefacts have reached a stage of well-established direct pipeline. However, the corresponding methods in II are in an evolving stage; but rapid advances are being made thanks to the emerging powerful methods of Machine Learning techniques. Although, Signal-to-Noise Ratio (SNR) in case of II compared to MSI has been a matter of concern in recent past, with availability of fast and high-time resolution photon counters and with upcoming facilities, this concern would be amply addressed in coming years.

In this article, we present the results of simulation of II observation of Polaris Aa (a Cepheid star in the Polaris system) and estimation of its radius parameters. We have two objectives of carrying out this study: 
\begin{itemize}
	\item{Validating II as a technique to observe and measure Cepheids}
	\item{Persuading the Indian astronomy community to adopt this versatile technique that can not only augment the current stellar measurement capabilities but can facilitate measurements of very high angular resolution at little or no additional establishment costs as indicated below.}
\end{itemize} 

In this context, we have simulated II observations of Polaris Aa using two different and separate hypothetical IACT facilities, and very similar to two such facilities currently operational in India and have used the data generated from these simulations to carry out Bayesian estimation of radius parameters of Polaris. The two operational gamma-ray facilities in India are considered as;
\begin{itemize}
	\item{a single large Imaging Atmospheric Cherenkov Telescope (IACT) set up similar to the Major Atmospheric Cherenkov Experiment Telescope (MACE) at Hanle, Ladakh}
	\item{and a subset array of three telescopes similar to the TeV Atmospheric Cherenkov Telescope with Imaging Camera (TACTIC) at Mount Abu, Rajasthan}.
\end{itemize}
These simulations are conducted as a demonstration of potential secondary science capabilities of these observatories \citep{yadav2009tacticmacegammaraytelescopes}. The following sections present the procedure and the key outcomes of this study. The section titled Polaris Observation with II presents our time-varying model of Polaris Aa, details the simulation of both the signal and noise in its II observation including the effects of Earth rotation during the observation. The subsequent two sections describe the II observations of Polaris Aa using salient observational parameters of MACE and TACTIC respectively. Finally, the Discussion section compares outcomes from both telescope configurations and discusses the potential role of such Cherenkov telescopes in advancing II-based astronomical studies in India.

%% file: polaris_observation.tex
\section{Polaris Observation with II}
Polaris, also known as Alpha Ursae Minoris ($\alpha$ UMi), is a triple star system located in the northern sky. The primary component, Polaris Aa, forms a close binary with Polaris Ab, separated by approximately 0.17 arc-seconds \citep{Evans_2008}. The third component, Polaris B, orbits this inner binary at a projected separation of about 18 arc-seconds \citep{roemer1965orbital}. Polaris Aa, commonly referred to as Polaris, is one of the nearest Cepheid variable stars and lies nearly aligned with Earth's rotational axis, appearing almost stationary in the night sky relative to other stars. Despite its proximity, several aspects of this star remain mysterious, particularly due to the uncertainties in its distance, which have led to discrepancies in the determination of its physical parameters \citep{van1995vizier, turner2012pulsation, nordgren2000astrophysical, merand2006extended}.

A striking characteristic of Polaris~Aa is its pulsating nature. The periodicity of its pulsation has been noted 3.97 days \citep{10.1093/mnras/stad2735}. Despite extensive study, the star's fluctuating period and its brightness variation - centered around an apparent magnitude of 2 in $V$ - band remain poorly understood. The variability of Polaris has been monitored since 1911, following its confirmation by Ejnar Hertzsprung \citep{hertzsprung1911nachweis}. However, the pulsation mode of Polaris remains a subject of debate, primarily due to its small and erratic amplitude variations over time \citep{usenko2018pulsational}. While some studies suggest that Polaris pulsates in the fundamental mode, others propose that it may be oscillating in its first or second overtone \citep{bono2001pulsation}. Recent observational studies have estimated the maximum radial variation of Polaris to be about 0.4 light-seconds (ls) \citep{2005AcA....55..247M}. In contrast, \cite{barron2026rotational} reported a larger value of approximately 0.93~ls. Motivated by these differing estimates, we adopt a radial variation model consisting of sinusoidal and cosinusoidal components with amplitudes of 0.8 and 0.9~ls, respectively (equation shown in upcoming subsection). This corresponds to a maximum radial variation of approximately 1.2~ls and is intended to represent the upper range of values for our simulation.

Accurately constraining its pulsational characteristics is crucial not only for the field of Asteroseismology but also for refining the calibration of the Cepheid period-luminosity relation, which is instrumental in extra-galactic distance measurements. These unresolved issues motivate the choice of Polaris Aa as the subject of this case study, particularly in the context of exploring its observability with Indian optical telescopes using the technique of II. 

We assume that Polaris Aa undergoes a simple harmonic radial pulsation following its fundamental mode, which we model using a simple trigonometric relation. The prior distributions required for the Bayesian estimation are generated using synthetic observation data, with hypothetical light collectors (or ``light buckets") having dimensions similar to those of MACE and TACTIC. The ground coordinates of these modeled mirrors are chosen to match the geographical locations of these two facilities.

\input{photon}
\input{snr}
\input{baseline}

%% file: photon.tex
\subsection{The Photon Flux from Model Polaris Aa}\label{sect:photflux}
Let the simple harmonic radial pulsation of the star with its radius $R(t)$ at time $t$ be given by 
\begin{equation}
	R(t) = R + R_1 \sin\left(\omega \, t\right) + R_2 \cos\left(\omega \, t\right)
	\label{eqn:radius}
\end{equation}
where $\omega = 2\pi/P$ is angular frequency of pulsation with periodicity $P$, $R$ is the mean radius, and $R_1$ and $R_2$ are the oscillatory coefficients in the variation of the mean radius.

It is needless to state that the true variability or pulsation of Cepheid stars is more intricate than a simple harmonic one. However, in the absence of real observational data which could be obtained by using telescopes, we have considered the variations to be of the simplest form. We wish to emphasize that potential to estimate the parameters of more complex functions associated with this variability exists.

Approximating the model Polaris Aa as a black body with an average uniform surface temperature $T$ located at a distance $D$ from the observatory and in the direction $\Omega$, the total spectral photon flux at frequency $\nu$ reaching the ground from the star at any time $t$ can be expressed as
\begin{equation}
	\Phi(\Omega, T) = |S(\Omega, T)|^2 \, \pi(R/D)^2 
	\label{eqn:flux} 
\end{equation}
where the photon flux per unit solid angle in one spectral channel $\nu$ coming from the direction $\Omega$ is given by
\begin{equation}
	|S(\Omega, T)|^2 = \frac{\displaystyle{\nu^2/c^2}}{\displaystyle{e^{h\nu/{kT(\Omega)}}} - 1}
	\label{spectral-phot-flux}
\end{equation}
giving the spectral photon flux per steradian from the star (with units of $\rm photons \, \rm m^{-2} \rm s^{-1} \rm sr^{-1} \rm Hz^{-1}$) reaching the ground from direction $\Omega$. $T(\Omega)$ in eqn.~(\ref{spectral-phot-flux}) represents the effective temperature map of the surface of the star with the average value being $T$.

\begin{figure}
	\includegraphics[width=\linewidth]{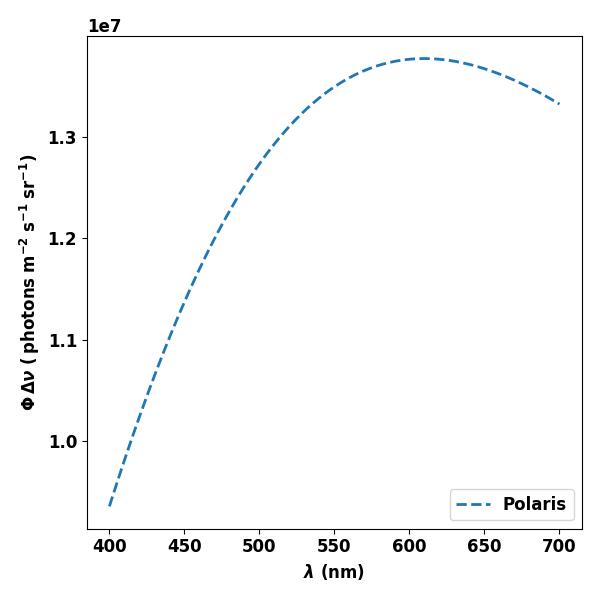}
	\caption{The photon flux from our model source Polaris Aa observed at mean observation wavelength $\lambda = 570 \,\rm nm$ and a bandwidth of photon detectors $\Delta \lambda = 1 \,\rm nm$.}
	\label{fig:fluxrate}
\end{figure}
\begin{figure*}
	\includegraphics[width=0.5\linewidth]{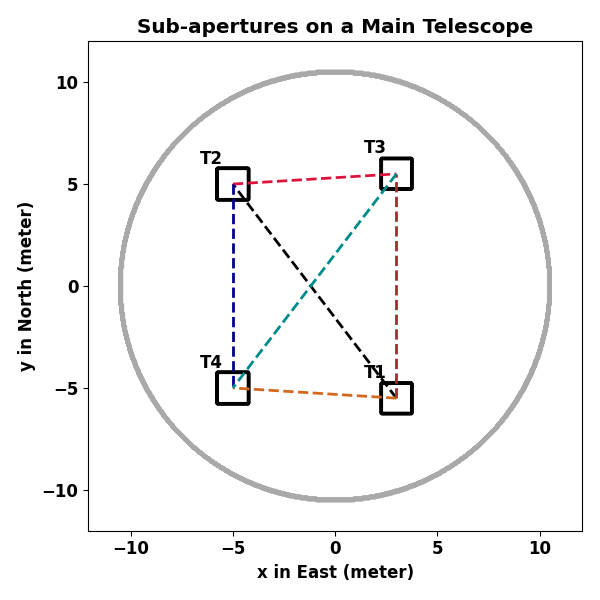}\hfil
	\includegraphics[width=0.5\linewidth]{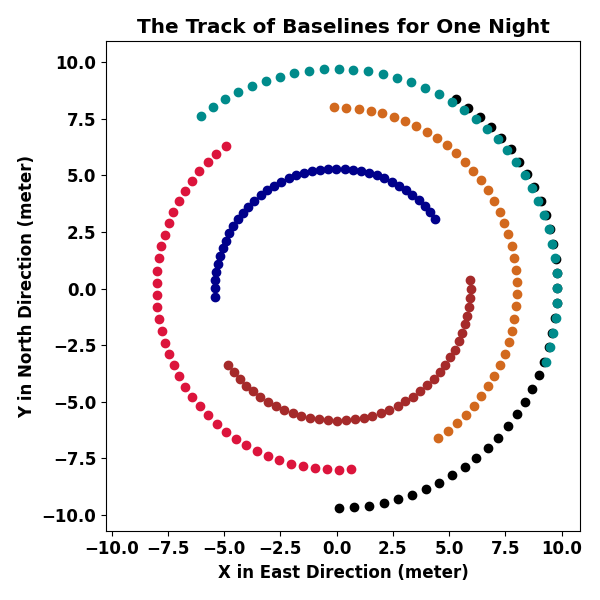}\hfil
	\caption{The left panel depicts a schematic representation of the four ``sub-apertures" (black-coloured rectangular boxes) configured on the objective (grey-coloured circle) of a MACE-like telescope. The detectors at the telescope's focus receive signals only from the mirror facets covered by these sub-apertures. The coloured dashed lines represent the six baselines provided by the four sub-apertures. In the first part of this work, this configuration is considered as an IACT facility to carry out simulated Intensity Interferometry (II) observations of the model Polaris Aa. The right panel illustrates the observational coverage during one night using these four sub-apertures. The color of each track corresponds to the coverage of the corresponding baseline shown on the left panel.}
	\label{fig:MACE}
\end{figure*}

\begin{figure*}
	\includegraphics[width=\linewidth]{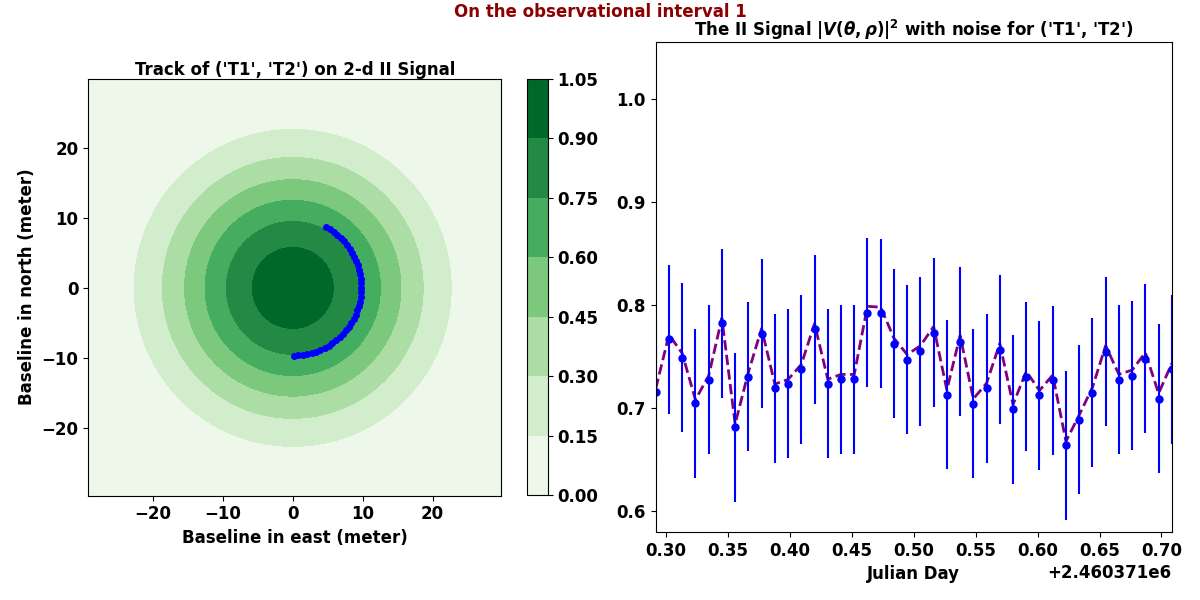}
	\includegraphics[width=\linewidth]{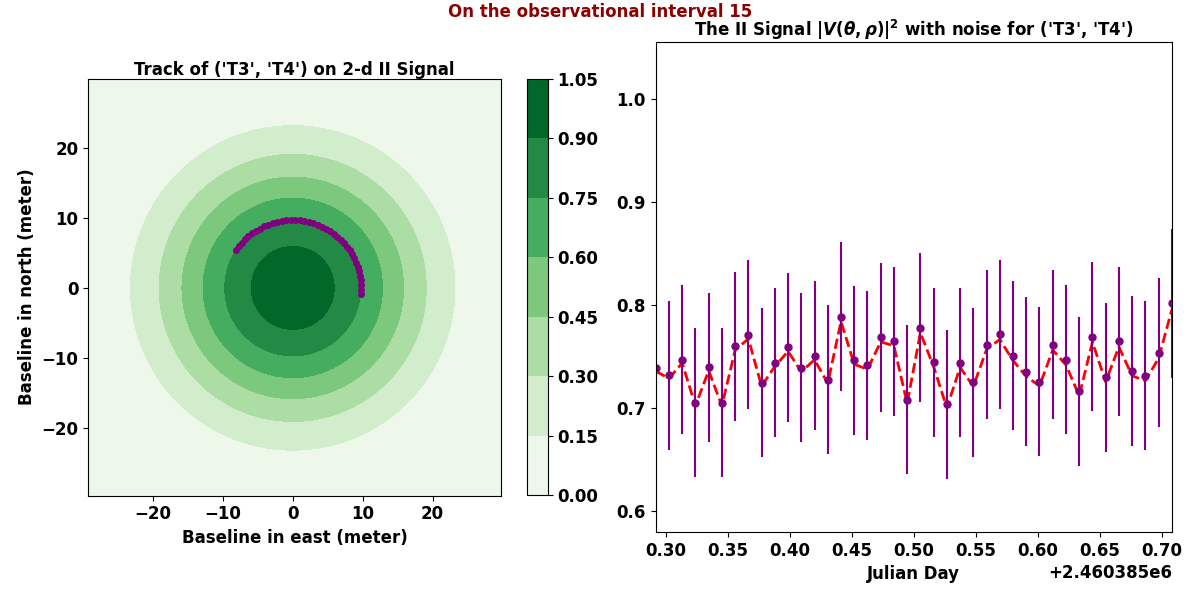}
	\caption{The left panels show the baseline tracks created from a single telescope with dense dotted lines on the observational plane, which is filled with the Airy disc of Polaris pulsating during two observational intervals - day 1 and day 15. The right panel corresponds to simulated II signals plotted against Julian Day. The points with error bars represents the signal assuming the oscillatory (pulsating) nature of Polaris, while the dashed line corresponds to a static (non-pulsating) source. The difference is small.  Also, since the data shown represents a single night's observation, the periodic variability - assumed to have a period of approximately four days - is not visible.}
	\label{fig:cntr_m}
\end{figure*}
\begin{figure}
	\includegraphics[width=0.97\linewidth]{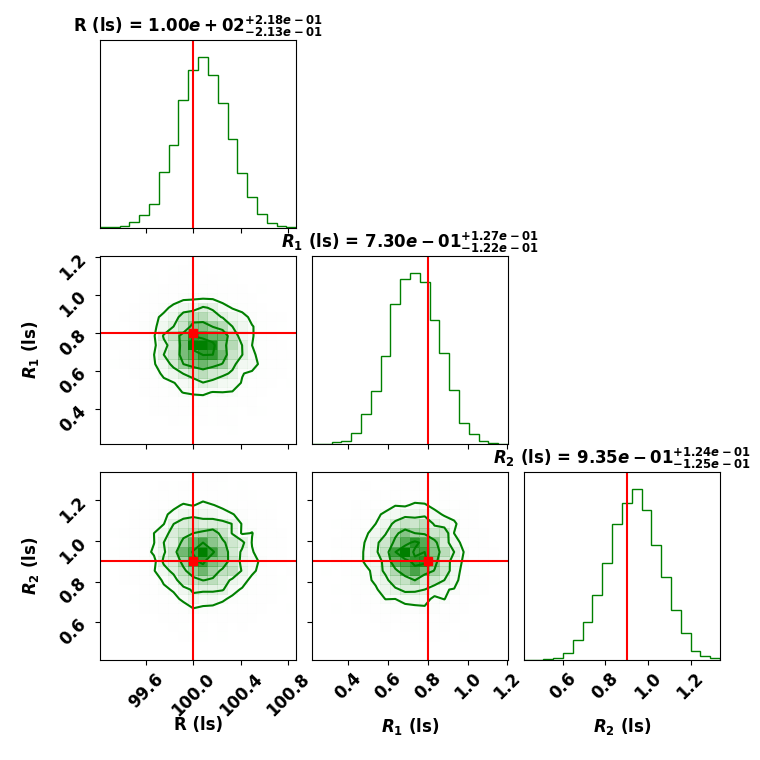}
	\caption{It shows the estimated mean radius and maximum radial variation of Polaris derived from simulated single telescope observations over 30 days each with 10 hours.}
	\label{fig:corner_m}
\end{figure}
\begin{figure*}
	\includegraphics[width=0.5\linewidth]{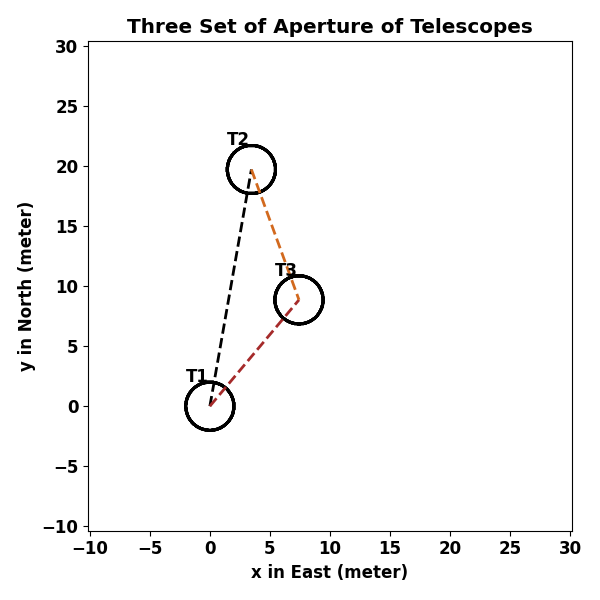}\hfil
	\includegraphics[width=0.5\linewidth]{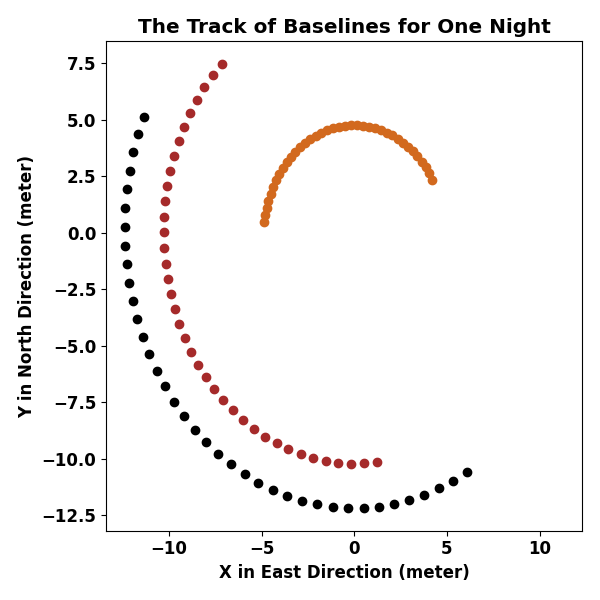}\hfil
	\caption{The left panel shows the positions of three selected telescopes array. The detectors used in these telescopes are photomultiplier tubes (PMTs). Right panel visualizes the observational coverage for a single night based on the baselines formed by the selected telescopes.}
	\label{fig:TACTIC}
\end{figure*}
\begin{figure*}
	\includegraphics[width=\linewidth]{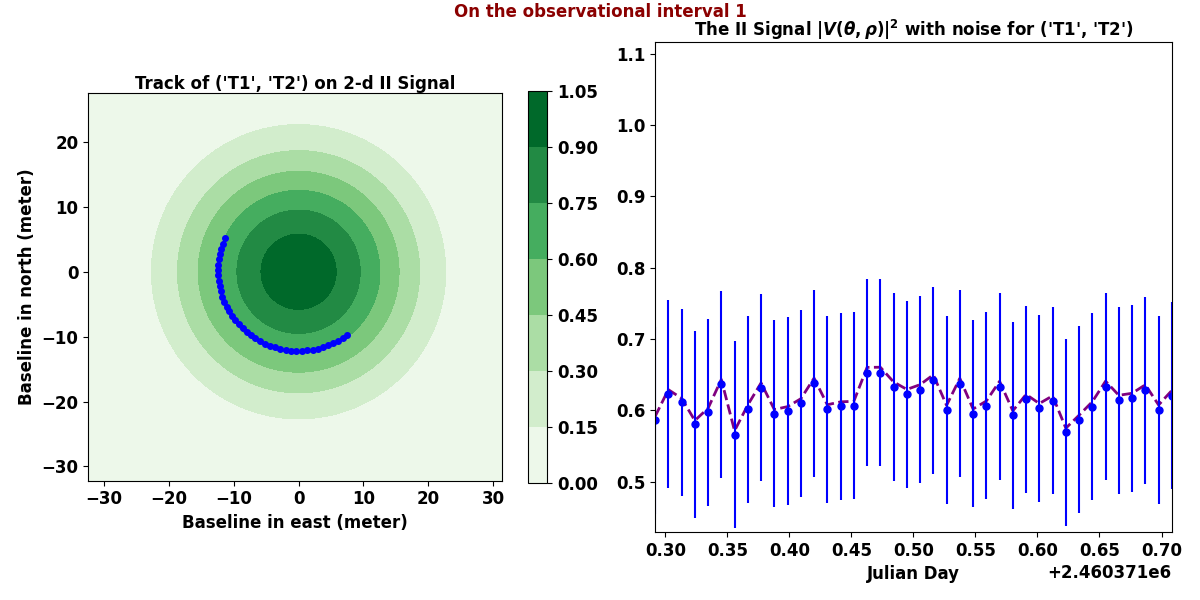}
	\includegraphics[width=\linewidth]{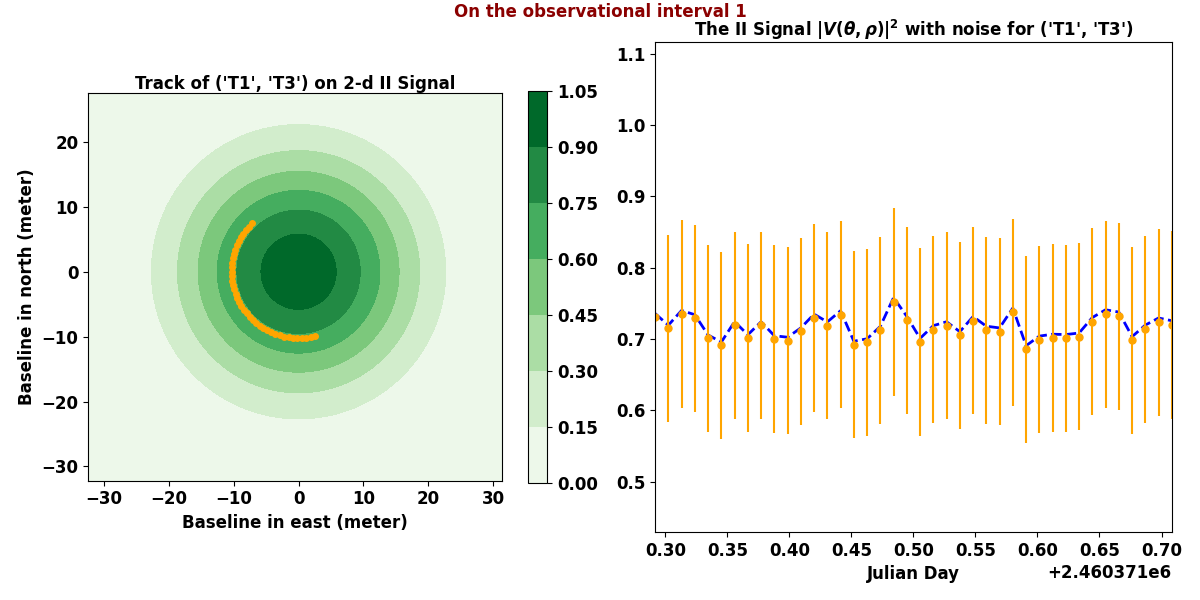}
	\caption{The left panels show the airy pattern of oscillatory Polaris. The blue and yellow densely dotted lines represent the tracking of the Airy pattern over one night of observation along the base lines $\rm{(T1,T2)}$ and $\rm{(T1,T3)}$ (as shown in Fig. (\ref{fig:TACTIC}). The right panels correspond to the simulated signals plotted against Julian Day. Two signal types are shown - one representing the oscillatory nature of Polaris (dotted lines with error bars) and the other representing a static model (dashed lines), as also depicted in Figure~\ref{fig:cntr_m}.}
	\label{fig:cntr_t}
\end{figure*}
\begin{figure}
	\includegraphics[width=\linewidth]{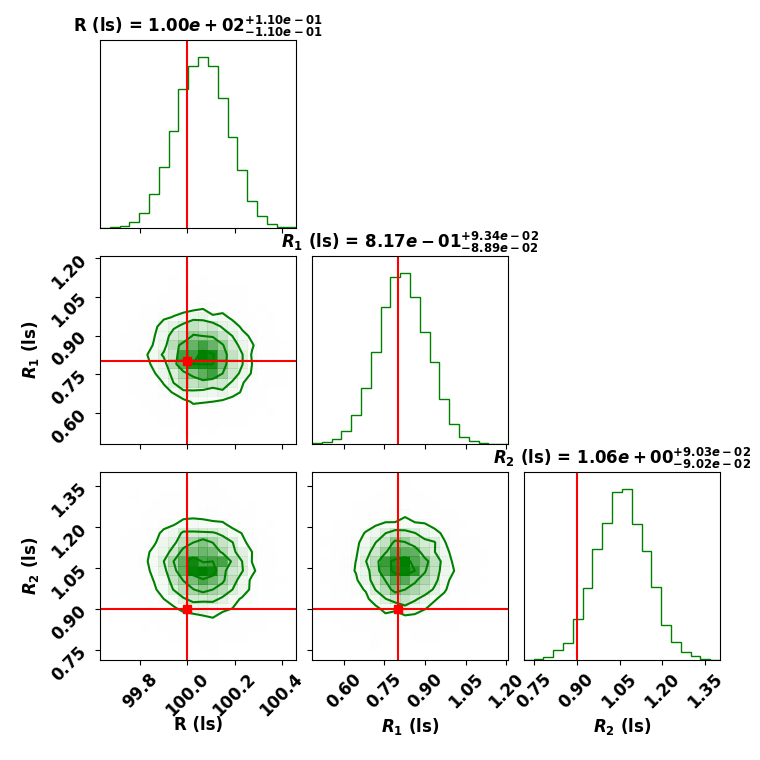}
	\caption{The estimated value of Polaris's mean and maximum variable radius with three telescope array.}
	\label{fig:corner_l}
\end{figure}
For a photon detector observing at frequency $\nu$ with bandwidth $\Delta \nu$, the photon flux coming from the source is
\begin{equation}
	\Phi(\Omega, T) \, \Delta \nu = \Phi(\Omega, T) \times \left|c\,\Delta \lambda/\lambda^2\right|
\end{equation}
where, $\lambda$ is the mean observing wavelength in the bandwidth of $\Delta \lambda$.

In his theoretical modeling of evolution of Polaris~Aa, \cite{10.1093/mnras/stv412} infers the distance from Earth $D$ and the radius $R$ to be 106~pc and $37.5R_{\odot}$ respectively. Recently published Gaia data releases and literature \citep{GaiaEDR3Summary2021, GaiaDR3Summary2023, Evans2024PolarisCHARA} have shown these values to be in the ranges  129–137~pc ($\sim 1.3\times 10^{10} - 1.4\times 10^{10}$ ls) and 43–53 $R_{\odot}$ ($\sim 100 - 123\,$ls) respectively. In our simulations reported here, we have adopted these values to be, respectively, $D = 1 \times 10^{10}\,$ls and the average radius of the source at time $t$ is taken to be $ R =  100\,$ls. Fig.~\ref{fig:fluxrate} shows the spectral distribution of photon flux reaching the ground from our model Polaris~Aa in the bandwidth $\Delta \lambda = 1 \, \rm nm$. For the mean observational wavelength $\lambda = 570 \, \rm nm$ and bandwidth $\Delta \lambda = 1 \, \rm nm$, the total photon flux is around $1.36 \times 10^7 \rm photons \,\, \rm m^{-2} \rm s^{-1}$, which is good photon statistics for II to resolve objects.

%% file: snr.tex
\subsection{The Signal and Noise for a Baseline}
The variation in radius of the star with time is accompanied by variation in its surface temperature; and, as a result the brightness (luminosity) of the star also varies. However, we assume an average surface temperature of the star during the variation cycle of its radius. The calculation of signal $S$ received by a pair of telescopes assumed to be located on an essentially horizontal plane and separated by a baseline vector ($\Delta_E, \Delta_N, \Delta_{up} \approx 0$) involves evaluation of the visibility signal $V(\theta,\rho)$. With the assumption of an average surface temperature of the star, the visibility signal on the plane of observation for a single star of radius $R$ and at a distance $D$ is given by
\begin{equation}
	V(\theta, \rho) = 2 \frac{J_1\left(\rho \, \theta \right)}{\left(\rho \, \theta \right)}
	\label{eqn:visib}
\end{equation}
where $\theta=R(t)/D$ is the angular size of star, the spatial frequency $\rho = 2 \pi \, \sqrt{u^2+v^2}$ with ($u = \Delta_E/\lambda$, $v= \Delta_N/\lambda$) and $J_1(\rho \, \theta)$ is the Bessel Function of the first kind providing the Airy disk intensity distribution due to the star on the ground.

Instead of measuring the complex visibility signal, II measures the squared magnitude $|V(\theta, \rho)|^2$ (or the ``squared visibility'') of the signal. As a consequence, the phase information is lost. However, the information about the shape and structure of the source can still be extracted from the squared visibility of the signal. From a practical point of view, we note that a pair of photon detectors fitted at the foci of their respective telescopes measure the intensities $I_1(\theta, \rho)$ and $I_2(\theta, \rho)$ simultaneously. The second-order correlation function
\begin{equation}
	g^{(2)}(\theta, \rho) = \frac{\langle I_1(\theta, \rho) \cdot I_2(\theta, \rho)\rangle}{\langle I_1\rangle \cdot \langle I_2 \rangle}
	\label{eq:g2-tau} 
\end{equation}
is computed by a correlator. Then the square magnitude of visibility function and the second order correlation function are related by the Siegert Relation (see e.g. Acciari et al \cite{acciari2020optical})
\begin{equation}
	g^{(2)}(\theta, \rho) - 1 = \frac{\Delta \tau}{\Delta t}|V(\theta, \rho)|^2.
	\label{eq:correl-visib} 
\end{equation}

It may be worthwhile to note the relationship between the expectation $\langle I \rangle$ of intensity $I$ (measured by an aperture--photon detector combination), the total photon flux $\Phi$ (according to eqn.~(\ref{eqn:flux})), and the effective aperture area $A$: $\langle I \rangle \propto A \Phi$. Both of these factors $\Phi$ and $A$ get canceled in the Siegert relation of eqn.~(\ref{eq:correl-visib}). However, the reasonably obvious relevance of these two quantities to the observed signal-to-noise ratio (SNR) is restored through the model of noise $N$ discussed a little hence.

We can now define the signal $S$ in terms of the ``HBT correlation", first suggested by Hanbury Brown and Twiss,
\begin{equation}
	S = g^{(2)}(\theta, \rho) = 1 + \frac{\Delta \tau}{\Delta t}|V(\theta, \rho)|^2
	\label{eqn:HBT}
\end{equation}
where $\Delta \tau$ is the coherence time of the observed radiation and $\Delta t$ is the temporal resolution of the photon detectors. $|V(\theta, \rho)|^2$, in this equation is used to fit the model of stellar structure. The information on $R(t)$ is extracted using Bayesian methods and the estimates of the mean radius and the parameters responsible for variations in it are inferred(as shown in \citep{10.1093/mnras/stab2391}).

An observed signal is always accompanied with noise. In fact, in a simulation, where noise also has to be modeled, it can be purposefully designed.  The noise $N$ that accompanies the signal $S$ collected from a telescope aperture of effective area $A$ in our simulation is 
\begin{equation}
	N = (A \, \Phi \, \eta \, n)^{-1} \sqrt{\frac{\Delta t}{t_{av}}} \, (\rm channels \times \rm length \, \rm of \, \rm observation)^{-1/2}
	\label{eqn:noise}
\end{equation}
where $\Phi$ is the total photon flux from the star (according to equ.\ref{eqn:flux}), $n$ is the loss in system, $\Delta t$ and $\eta$ are the photon detector's resolution time and its quantum efficiency respectively. $t_{av}$ is the average observational time over which the the raw time-series data $I_1(\theta, \rho)$ and $I_2(\theta, \rho)$ are smoothed by taking running-averages. The randomness in the noise $N$ ( which may arise from several factors such as system loss $n$, photon flux $\Phi$ etc.) is factored in through Gaussian sampling. According to eqn.~\ref{eqn:noise}, the noise decreases with the number of channels used in observation and the length of observation (ref Fig.9 of \cite{10.1093/mnras/stac2433} for illustration of this point). On this ground and with a view to obtaining a good signal-to-noise ratio (SNR) for robust estimations of the complex parameters of of Polaris Aa, we extend the simulation for a longer observational time, typically spanning several hours each night and for several nights. Two primary factors that contribute towards variations in the visibility signal  (equation~\ref{eqn:visib}) for Polaris-type variables are:
\begin{itemize}
	\item The change in angular size $\theta$ of the star, as described by equation~\ref{eqn:radius}.
	\item and the Earth's rotation, which plays a significant role in the time-variation of the projection of the baseline $(u, v)$ concerning the source. 
\end{itemize}

We next discuss the effects of the Earth's rotation on the baseline.

%% file: baseline.tex
\subsection{Rotation of Baseline}
The simulated signal from any stellar source observed over long hours overnight has to be modulated with the effect of the rotation of the Earth. This modulation is built into the interferometric coordinates $(u,v,w)$ corresponding to the fixed baseline vector $(\rm \Delta_E, \rm \Delta_N, \rm \Delta_{up})$ associated with a pair of telescopes on the ground as mentioned earlier, using three rotation matrices $\cal R$ as shown in eq.(\ref{eq:uvw-rotation}):
\begin{equation}
	\begin{pmatrix} u \\ v \\ w \end{pmatrix} = \frac 1\lambda \,
	{\cal R}_1(\delta) \, {\cal R}_2(h) \, {\cal R}_1(-l)
	\begin{pmatrix}
		\Delta_{\scriptscriptstyle{\rm E}} \\
		\Delta_{\scriptscriptstyle{\rm N}} \\
		\Delta_{\rm up}
	\end{pmatrix}
	.
	\label{eq:uvw-rotation}
\end{equation}
Here, $(u, v)$ is the baseline vector components in the east and north directions, and $w$ is the perpendicular coordinate along the line of sight to the star (Polaris Aa). The rotation matrices are constructed with the latitude $l$ of the setup, the declination $\delta$ and the hour angle $h$ of the source (Polaris Aa). The individual rotation matrices have the usual form
\begin{equation}
	{\cal R}_1(f) =
	\begin{pmatrix}
		1 & 0 & 0 \\
		0 & \cos f  & -\sin f \\
		0 & \sin f  &  \cos f
	\end{pmatrix},
	\label{eqn:R1}
\end{equation}
${\cal R}_2$ and ${\cal R}_3$ following by a cyclic permutation of indices.

In the upcoming section, we will see that these three matrices will be applied according to the position of the telescope with respect to the source. It is to be noted that, the latitude $l$ of the observatory and declination $\delta$ of the observed source being constant, the time variation in the $(u,v,w)$ coordinates results from the time variation $h(t)$ of the hour-angle of the source during the observation.

The above applies if the individual apertures stay fixed on the ground.  If the apertures are sub-apertures on a dish that is slewing to track the target, the above gets modified.  Consider a composite rotation
\begin{equation} \label{eq:altaz}
  {\cal R}_3(\psi) \, {\cal R}_1(\theta) \, {\cal R}_3(\phi)
\end{equation}
equal to the composite rotation
\begin{equation}
  {\cal R}_1(\delta) \, {\cal R}_2(h) \, {\cal R}_1(-l)
\end{equation}
from Eq.~(\ref{eq:uvw-rotation}), but made up of different subrotations around different axes.  That is, we have a rotation in two different Euler-angle conventions, and standard formulas for changing between conventions give $\psi\,\theta\,\phi$ in terms of $\delta,h,l$.  While the composite rotation is the same in both cases, the individual angles have different geometric meanings.  Of the new set of angles $\phi$ is an azimuthal rotation, $\pi/2-\theta$ is an altitude, and $\psi$ is the field rotation.  If the apertures are on a dish that slews along altitude and azimuth, the $\theta$ and $\phi$ contributions are compensated for by the slew, leaving only the field rotation
\begin{equation}
	\begin{pmatrix} u \\ v \\ w \end{pmatrix} = \frac 1\lambda \,
	{\cal R}_3(\psi)
	\begin{pmatrix}
		\Delta_{\scriptscriptstyle{\rm E}} \\
		\Delta_{\scriptscriptstyle{\rm N}} \\
		\Delta_{\rm up}
	\end{pmatrix} .
	\label{eq:altaz-rotation}
\end{equation}
For Polaris there is not much difference between the two cases, since the telescope does not move much while tracking the star.  Hence in our simulations we will use Eq.~(\ref{eq:uvw-rotation}) throughout.

\subsection{Prior, Likelihood and Estimation of Parameters}
The time-dependent simulated intensity interferometry (II) data, generated using the baseline rotation formalism of Eq.~(\ref{eq:uvw-rotation}), are analysed within a Bayesian inference framework. The goal is to infer the posterior probability distributions of the parameters that describe the radial pulsation of Polaris~Aa. For efficient posterior sampling and evidence estimation we employ the nested sampling package \texttt{dynesty} \citep{speagle2020dynesty}.

The model is fully specified by the three parameters collected in the vector
\[
\Theta = (R, R_1, R_2),
\]
where \(R\) is the mean stellar radius and \(R_1\), \(R_2\) are the amplitudes of the sinusoidal and cosinusoidal terms in the radial pulsation model (Eq.~\ref{eqn:radius} of Sect.\ref{sect:photflux}). The angular frequency is fixed at the observed pulsation period of Polaris~Aa, \(\omega = 2\pi/P\) with \(P \approx 3.97\,\text{d}\).

We adopt uniform priors over ranges that comfortably encompass literature values while remaining physically plausible:
\begin{equation}
  \begin{aligned}
    p_{\min} &= [0.1\,R^{c},\, 0.1\,R^{c}_1,\, 0.1\,R^{c}_2], \\
    p_{\max} &= [1.8\,R^{c},\, 1.8\,R^{c}_1,\, 1.8\,R^{c}_2].
  \end{aligned}
  \label{eqn:prior}
\end{equation}
As discussed in Sect.~\ref{sect:photflux}, we choose the ground-truth value of the average radius ($R^{c} = 100\,$ls). 
The ground-truth values for the pulsation amplitudes are chosen as (\(R^{c}_1 = 0.8\,\text{ls}\), \(R^{c}_2 = 0.9\,\text{ls}\)) following \cite{barron2026rotational}.
These values are indicated by the red lines in the corner plots (Figs.~\ref{fig:corner_m} and~\ref{fig:corner_l}).

The likelihood function compares the simulated correlation measurements $|V_s|^2$ (including realistic noise and Earth-rotation effects) with the model predictions $|V_m|^2$. We adopt a Gaussian form with an additional nuisance scaling parameter \(\alpha\) that absorbs possible overall calibration offsets in the measured correlation function:
\begin{equation}
	\ln \mathcal{L}\!\left(|V_s|^2 \middle| \alpha, \Theta\right)
	= -\frac12 \sum_s
	\frac{\bigl[|V_s|^2 - \alpha\, |V_m|^2(\Theta)\bigr]^2}{\sigma_s^2}.
	\label{eqn:likelihood}
\end{equation}
Here \(|V_m|^{(2)}(\Theta)\) is the squared visibility (Eq.~(\ref{eqn:visib})) at the instantaneous angular size \(\theta(t) = R(t)/D\) and the time-dependent baseline \(\rho(t)\) that incorporates Earth rotation. The parameter \(\alpha\) is marginalised over analytically during the sampling, following the procedure described in Sect.~4.2 of \cite{10.1093/mnras/stab2391}.

The posterior probability distribution follows directly from Bayes’ theorem:
\begin{equation}
	P\!\left(\Theta \middle| |V_s|^{(2)}\right)
	\propto
	\mathcal{L}\!\left(|V_s|^{(2)} \middle| \Theta\right)\,\pi(\Theta),
	\label{eqn:posterior}
\end{equation}
where \(\pi(\Theta)\) denotes the prior (Eq.~\ref{eqn:prior}). The \texttt{dynesty} sampler returns a set of posterior samples together with an estimate of the Bayesian evidence. These samples are used to construct the marginalised one- and two-dimensional posterior distributions for \(R\), \(R_1\) and \(R_2\), from which we extract best-fit values and credible intervals (shown in the corner plots of Figs.~\ref{fig:corner_m} and~\ref{fig:corner_l}).

This Bayesian framework is applied in the following sections to two distinct observational configurations: a single large Cherenkov telescope similar to MACE employing the split-mirror technique and a three-telescope array similar to TACTIC.

%% file: mace.tex
\section{II Simulation with MACE-like Single Large IACT}
The Major Atmospheric Cherenkov Experiment (MACE) telescope is an advanced gamma-ray telescope located in Hanle, Ladakh, India at an altitude of 4270 meters above the sea level. It features a 21-meter diameter reflector composed of 1424 mirror facets, each measuring approximately 0.5 m × 0.5 m in size \citep{bhatt2021status}. The high-altitude location, large aperture, and optical quality make it a promising candidate for applications in II. However, with only a single telescope, one is limited to measuring zero-baseline intensity correlations, and this restricts spatial resolution. Multiple baselines are essential to capture interference fringes to retrieve detailed information about the source structure and morphology. Here, we wish to point out that the substantial aperture of a MACE-like telescope objective with tessellated mirror facets offers possibilities of implementing a split-mirror technique and emulating multiple baselines, similar to other recent proposals \citep{2025MNRAS.538..867M,2025PASP..137l5004C}.  This can be implemented by configuring a set of sub-apertures at fixed distances with respect to each other and spread across the tessellated objective. Each such sub-aperture, possibly composed  of a group of contiguous mirror facets along with their respective back-end optics and electronics, acts as a small IACT telescope. Using groups of such sub-apertures now acting as independent but co-relatable IACTS and combined with Earth rotation, would be able to emulate multiple baselines. With such a configuration of sub-apertures on a MACE-like telescope, II observations using HBT correlations of signals from stellar sources to obtain high angular resolutions can be achieved. We discuss the simulation of the II observation of the model Polaris Aa with implementation of the split-mirror technique on the telescope below.

\begin{table*}
	\centering
	\begin{tabular}{lccccccccc}
		\toprule
		\textbf{Configuration} &
		\textbf{${\omega}\;(\mathrm{rad\,s^{-1}})$} &
		\textbf{Night (Hour)} &
		\textbf{$A_{\mathrm{eff}}\;(\mathrm{m^2})$} &
		\textbf{Channel} &
		\textbf{$\sigma_s$} &
		\textbf{SNR} &
		\textbf{$R$ (ls)} &
		\textbf{$R_1$ (ls)} &
		\textbf{$R_2$ (ls)} \\
		\midrule
		
		{Split-Mirror}
		& $1.8e-5$ & 30 (10 hours) & 2 & 2 & 39.85 & 2212 & $100 \pm 0.22$ & $0.73 \pm 0.13$ & $0.93 \pm 0.12$ \\
		[2mm]
		{3 IACTs Array}
		& $1.8e-5$ & 30 (10 hours) & 2 & 4 & 28.18 & 2085 & $100 \pm 0.11$ & $0.82 \pm 0.09$ & $1.06 \pm 0.09$ \\
		
		\bottomrule
	\end{tabular}
	\caption{Comparison of the estimated stellar parameters obtained using the split-mirror technique on single telescope and a three-IACT array for the adopted observational configurations.}
	\label{table:para}
\end{table*}

\subsection{Split Mirror Technique with MACE-like Telescope}
In this study, we consider an observational configuration in which a block of nine neighboring mirror facets, each with approximately identical focal lengths, is grouped into a square panel measuring 1.5 m × 1.5 m. These panels represent a modular sub-aperture unit of the telescope. As illustrated in the left panel of Figure~\ref{fig:MACE}, our simulation utilizes the signal collected from four such mirror panels (represented by the four small square shapes) distributed at distinct positions on the circular primary reflector (circular shape in grey color) of the telescope.
In this configuration, the photo-detector assembly of the telescope's imaging camera is restricted to receiving light exclusively from these four spatially separated sub-apertures. This method, known as the split-mirror technique, enables the creation of multiple virtual baselines, which are essential for II. These virtual baselines mimic the effect of physically separated telescopes, thereby enhancing the spatial sampling of the source's brightness distribution. It may be noted that the lengths of these baselines among the sub-apertures range between $5.31\, \mathrm{m}$ to $9.93\, \mathrm{m}$.
The right panel of Figure~\ref{fig:MACE} depicts the tracks traced by the six possible baselines formed due to Earth's rotation between the four panels during a single night of observation. While this figure shows the coverage for one night, multiple nights of observation are typically accumulated to improve the signal-to-noise ratio (SNR) and achieve reliable II measurements.

The noise associated with the simulated signal is computed using the expression defined in Equation~\ref{eqn:noise}. This expression incorporates contributions from the total photon flux, instrumental characteristics of the telescope, number of channels, length of observation and other observational parameters. For calculating the photon flux, Polaris Aa is assumed to have an average surface temperature of 6015 K. The resolution time of the photon detectors is taken as $\Delta t = 1 \, \rm ns$, and each observation session spans 15 minutes. The simulation assumes 2 detection channel \citep{vogel2025simultaneous} and a total observational duration of 30 nights each for 10 hours. In this simulation, the quantum efficiency of the photon detector ($\eta$) and the signal transmission factor ($n$) are assumed to be unity, i.e., $\eta = 1$ and $n = 1$. These instrumental losses are instead incorporated through the effective light-collecting area. Each square sub-aperture has a geometric area of $1.5\,\mathrm{m} \times 1.5\,\mathrm{m} = 2.25\,\mathrm{m}^{2}$. However, an effective collecting area of $A = 2\,\mathrm{m}^{2}$ is adopted in the simulations, corresponding to $88.9\%$ of the geometric area. The remaining $11.1\%$ is assumed to account for photon losses arising from finite optical throughput, mirror reflectivity, detector inefficiencies, and other instrumental effects. This treatment provides a simplified representation of instrumental losses and should therefore be regarded as an idealized approximation. Nevertheless, these effects are expected to influence the achievable signal-to-noise ratio quantitatively rather than alter the qualitative feasibility conclusions presented in this work.

Under these conditions, the total signal-to-noise ratio (SNR) achieved over 30 nights of observation is of the order of 2200. This SNR is sufficiently high to reliably estimate both the average radius and its variation for Polaris Aa. Notably, the determination of only the mean radius does not require such a high SNR, implying that even shorter observation durations could be acceptable for certain objectives.

\subsection{The Signal and Estimation of Parameters}
Figure~\ref{fig:cntr_m} presents the results of simulated observations for a single night. The left panels display the contour maps of the HBT correlation signal over a specific time interval (averaged over a few minutes). Superimposed on these maps are the baseline tracks of the telescope pairs (T1, T2) and (T3, T4), shown in the upper-left and lower-left panels, respectively. These tracks, represented by dense dotted lines, illustrate the evolution of the projected baselines due to Earth's rotation.

The right panel of the figure shows the correlated signal as a function of Julian Day for the tracked baselines over the observation of one night only. Two simulated signals are presented: one assuming Polaris Aa exhibits oscillatory radial variation (plotted as pointed lines with error bars) and the other assuming a static nature of the star (shown as a dashed line). Since the observation is limited to one night, the periodic pattern corresponding to Polaris Aa’s pulsation, assumed to have a period of four days, is not yet visible in the data.

To estimate stellar parameters, we simulate a longer observational campaign spanning 30 nights. Using the nested sampling algorithm implemented in \texttt{dynesty} \citep{speagle2020dynesty}, we infer the oscillatory characteristics of Polaris Aa under the assumption of radial pulsation. The results are presented in Figure~\ref{fig:corner_m}. With a sufficiently high signal-to-noise ratio achievable using a single large telescope, the mean stellar radius ($R$) and the amplitudes of its oscillation ($R_1$ and $R_2$) are estimated with uncertainties of approximately $0.22 \%$, $17.39 \%$, and $13.26 \%$, respectively. This successful estimation of the three parameters strongly suggests that real-time observation be mounted using the MACE telescope. Such observations would enable testing of more complex oscillation models of Polaris Aa and other Cepheids. 

%% file: tactic.tex
\section{II Simulation With An Array of three-IACTs}
Here, we present the details of the results of oscillatory parameter estimation of model Polaris Aa using simulated II observation with a three-IACT array partially similar to the IACT configuration of the TACTIC facility at Mount Abu, Rajasthan, India. The TeV Atmospheric Cherenkov Telescope with Imaging Camera (TACTIC) represents India's first imaging atmospheric Cherenkov telescope (IACT) dedicated to very high-energy (VHE) gamma-ray astronomy. The TACTIC array comprises four telescopes: three positioned at the vertices of an equilateral triangle with a side length of approximately 20 meters and one located at its centroid \citep{tickoo1999drive}. 

In our simulation, we have considered only three IACTs at the vertices of an equilateral triangle while maintaining other parameters like the geometric configuration and instrumental precision similar to those of TACTIC. For example, similar to TACTIC, we also consider each telescope to be constructed using 34 circular mirror facets, each 0.6 m in diameter, yielding a total effective light-collecting area of approximately $9.5 \rm m^2$ per telescope \citep{koul2007tactic}. Besides, each telescope is equipped with a photo-multiplier tube (PMT) featuring a time resolution of 3 ns. This arrangement of the telescopes ensures a well-distributed coverage across the observational plane, enabling effective photon correlation measurements.

\subsection{SNR estimation with the Three-IACT array}
The left panel of Figure~\ref{fig:TACTIC} represents the geometrical telescope configuration considered in this simulation. The right panel of the figure illustrates the tracking of the three baselines formed by this configuration, plotted in the same format as the case of the Split-mirror configuration adopted for a MACE-like telescope in Figure~\ref{fig:MACE}. The instrumental characteristics of these telescopes are incorporated into the signal simulation through a reduced effective area $A$ in eqn.(\ref{eqn:noise}). 
Only $20\%$ of each aperture's area is considered to account for system inefficiencies such as quantum efficiency and signal loss, resulting in an effective photon-collecting area of approximately $2 \rm m^2$ per telescope. While the mean radius of Polaris Aa can be inferred from a shorter observation period, accurately estimating the maximum radius variation requires a high signal-to-noise ratio (SNR), so a long observational period is needed. To achieve this, we simulate 30 nights each with 10 hours of observation using 4 channels, yielding an SNR of approximately 2085. An equivalent SNR could also be obtained by observing for 120 nights with a single channel. Alternatively, the required observing time could be significantly reduced by increasing the effective photon-collecting area through improved detector quantum efficiency, reduced optical losses, or a larger fraction of the telescope aperture contributing to the detected signal. The simulated II signal space with two (out of the possible three) baselines are presented in Figure~\ref{fig:cntr_t} for a single night of observation. The left panels show the simulated II signal as a green contour map with the signal strength becoming weaker with increasing baseline. The actual coverage of this signal space by the two baselines with the tracking of baselines marked by the dense and coloured dotted lines are also shown in these left panels. The right panels present the extracted signal from these tracks, highlighting the distinction between Polaris Aa's oscillatory (pointed with error bars) and static (dashed line) models. 

\subsection{The Parameter Estimation with Three Arrays}
The signal strength depicted in the right panel of Figure~\ref{fig:cntr_t} demonstrates sufficient quality to recover the simple trigonometric parameters associated with the pulsating model of Polaris Aa (Eqn.~\ref{eqn:radius}). Using simulated data from the three-IACT configuration, we have estimated the mean radius $R$ and the amplitudes $R1$ and $R2$ of its variation. The results, presented in Figure~\ref{fig:corner_l}, indicate that the mean radius $R$ of Polaris Aa is determined with an uncertainty of approximately $0.11 \%$. The oscillatory parameters, $R_1$ and $R_2$ are estimated with $11.43 \%$ and $8.51 \%$ uncertainty respectively. Notably, these uncertainties are lower than those obtained using the MACE-like single telescope simulation (see Figure~\ref{fig:corner_m}), suggesting improved parameter constraints with the three-IACT TACTIC-like configuration.

%% file: summary.tex
\section{Summary and Comparison of Results}

We have carried out end-to-end simulations of intensity interferometry observations of the pulsating star Polaris Aa using two observational configurations modelled on existing Indian Cherenkov facilities. The principal results from the two campaigns are collected in Table~\ref{table:para} for direct comparison.

Both configurations achieve high signal-to-noise ratios (SNR $\gtrsim 2000$) after 30 nights of 10-hour observations, sufficient to recover the mean stellar radius $R$ and the amplitudes of its radial pulsation ($R_1$ and $R_2$) with useful precision. In the single-telescope configuration employing the split-mirror technique (Section~3), six virtual baselines are generated from four sub-apertures. With two detection channels the simulation yields an SNR of 2200 and returns the parameters
\[
R = 100 \pm 0.22\,\text{ls},\,
R_1 = 0.73 \pm 0.13\,\text{ls},\,
R_2 = 0.93 \pm 0.12\,\text{ls}.
\]

The fractional uncertainties are approximately 0.22 per cent on the mean radius and 18--13 per cent on the oscillatory amplitudes.

In the three-telescope array configuration (Section~4), only three physical baselines are available, but the use of four detection channels improves the noise properties. The resulting SNR is 2085 and the recovered parameters are
\[
R = 100 \pm 0.11\,\text{ls},\,
R_1 = 0.82 \pm 0.09\,\text{ls},\,
R_2 = 1.06 \pm 0.09\,\text{ls},
\]
corresponding to fractional uncertainties of 0.11 per cent on $R$ and 11--8 per cent on the pulsation amplitudes.

The comparison in Table~\ref{table:para} shows that the three-telescope array delivers modestly tighter constraints on the oscillatory parameters despite having fewer baselines. This improvement is attributable to the larger number of detection channels employed. Conversely, the single-telescope split-mirror setup benefits from denser $(u,v)$ coverage (six baselines versus three), which may become advantageous when more complex source morphologies are investigated. In both cases the mean radius is recovered to sub-percent accuracy, demonstrating that even a modest observational campaign with existing Indian Cherenkov telescopes can yield scientifically useful measurements of Cepheid pulsation parameters via intensity interferometry.

These quantitative outcomes confirm that the two facilities can serve as effective platforms for II observations of bright, variable stars and provide a concrete benchmark for planning future real observations.

%% file: conclusion.tex
\section{Discussion and Outlook}
The resurgence of II as a technique for high-resolution stellar imaging has been significantly driven by its adoption during moonlit nights by several different Imaging Atmospheric Cherenkov Telescopes (IACT) arrays. The combination of large collecting apertures and long baselines makes these facilities particularly well-suited for II, offering the signal-to-noise ratios necessary to resolve fine details in stellar structures at optical wavelengths. Recent advancements by the VERITAS, MAGIC, and HESS arrays have demonstrated the feasibility of using II for the angular resolution of complex stellar systems, including OB-type stars, Wolf-Rayet stars, and pulsating stars in binaries and multiple systems \citep{acharyya2024angular, abe2024performance, vogel2025simultaneous, 10.1093/mnras/stac1617}.

In this work, we have explored the secondary science potential of Indian Cherenkov facilities - specifically similar to the MACE and the TACTIC observatory - for II observations. Using a simplified model of Polaris Aa as a test case, we have successfully simulated two separate II observation campaigns using two virtual IACT configurations respectively very similar to these telescopes and have estimated its mean radius and oscillatory behavior with good precision. The results indicate that even with a limited number of baselines and modest instrumental configurations, valuable astrophysical parameters can be extracted. 

The results presented here should, however, be regarded as idealized feasibility estimates. Polaris Aa is modeled as an isolated star with a uniform surface temperature undergoing simple sinusoidal radial pulsations. The simulations neglect several physical and instrumental effects, including possible tidal effects from the known companions of Polaris Aa, limb darkening, temperature variations, calibration systematics, and other instrumental non-idealities. Instrumental losses are represented through the adopted effective collecting area rather than by explicitly modeling detector efficiency and optical throughput. These simplifications are expected primarily to affect the achievable signal-to-noise ratio and the precision of the recovered parameters, but they are not expected to alter the qualitative conclusion that Indian Cherenkov telescopes also can be effectively utilized for Intensity Interferometric studies of bright pulsating stars such as Polaris Aa.

These findings not only highlight the feasibility of utilizing Indian IACTs for II during non-gamma observational periods, such as moonlit nights, but also strongly suggest that real-time collaborative efforts be undertaken using these existing Indian facilities. At very little or no additional costs, Indian marks would be made in this prospective and resurging field of Astronomy. Moreover, the success of this approach points toward a broader potential: other optical telescopes with sensitive photon detectors at the focus may also be adapted for II applications. Looking forward, the extension of this method beyond the northern sky, combined with collaborative efforts involving southern hemisphere facilities, promises to open a new dimension in high-resolution Indian optical astronomy.